\documentclass[aps,prl,showpacs,amsmath,amsfonts,superscriptaddress,twocolumn]{revtex4}

\usepackage{graphicx}
\usepackage{dcolumn}
\usepackage{bm}
\usepackage{color}

\begin{document}

\author{Lode Pollet}
\affiliation{Department of Physics, Harvard University, Cambridge, Massachusetts 02138, USA}
\author{Nikolay V. Prokof'ev}
\affiliation{Department of Physics, University of Massachusetts,
Amherst, MA 01003, USA}
\affiliation{Russian Research Center ``Kurchatov Institute'',
123182 Moscow, Russia}
\author{Boris V. Svistunov}
\affiliation{Department of Physics, University of Massachusetts,
Amherst, MA 01003, USA}
\affiliation{Russian Research Center ``Kurchatov Institute'',
123182 Moscow, Russia}

\title{Regularization of Diagrammatic Series with Zero Convergence Radius}

\begin{abstract}
The divergence of perturbative expansions for vast majority of macroscopic systems,
which follows from Dyson's collapse argument, prevents Feynman's diagrammatic
technique from being directly used for controllable studies of strongly
interacting systems. We show how the problem of divergence can be solved
by replacing the original model with a convergent sequence of successive
approximations which have a convergent perturbative series.
As a prototypical model, we consider the zero-dimensional $\vert \psi \vert^4$ theory.
\end{abstract}



\date{\today}

\maketitle

Almost sixty years ago, Dyson provided a physical argument why power-series expansions
in quantum electrodynamics are divergent even after proper charge
and mass renormalization \cite{Dyson52}.
The argument is straightforwardly generalizable to a wide range of
classical and quantum statistical problems and has far-reaching
consequences for any perturbative approach~\cite{NegeleOrland}.
It goes as follows: The interaction potential between two electrons with charge $e$
is of the form $e^2/r_{12}$, with $r_{12}$ the distance between
them. When performing a series expansion in
$e^2$ ({\it i.e.}, the interaction strength) a physical
quantity of interest can be expressed as
\begin{equation}
F(e^2) = a_0 + a_2e^2 + a_4 e^4 + \ldots
\label{eq:QED}
\end{equation}
If the series have finite convergence radius $\xi$ then $F(e^2)$
is an analytic function at $e=0$, implying that for sufficiently
small values of $e$, $F(-e^2)$ is a well-behaved analytic function.
However, $F(-e^2)$ corresponds to a fictitious world with purely
imaginary particle charges where the interaction potential is
of the form $-e^2/r_{12}$, and is thus attractive.
In this fictitious world, the vacuum state would be unstable
against production of an infinite number of electron-positron pairs,
each put in a separate region of space since the gain
in the negative Coulomb energy is larger than the increase in the kinetic energy.
Hence, Dyson concluded that $F(e^2)$ cannot be analytic around $e=0$ and that the
convergence radius must be zero, $\xi = 0$. \\

The $\vert \psi \vert^4$ theory is associated with the following
partition function in $d$ dimensions:
\begin{equation}
Z = \int {\mathcal D}\psi \,    {\rm e}^{- \! \int \! d^dr \, \left\{  \vert \nabla \psi \vert^2 + \lambda \vert \psi \vert^4 \right\} } ,
\label{eq:psi_four}
\end{equation}
describing the statistics of the classical complex-valued field $\psi({\bf r})$.
This effective field theory is often used to describe critical behavior of the
superfluid phase transitions in interacting Bose systems.
The model (\ref{eq:psi_four}) can be solved numerically very efficiently
by a number of methods, but none of them is based on an expansion
in $\lambda$. A successful regularization of the perturbative expansion
in $\lambda $ for the $\vert \psi \vert^4$ theory would change the
status of the standard diagrammatic technique to that of a
systematic method for accurate studies of strongly correlated systems,
and provide the basis for numerical treatments in the framework of (bold)
Diagrammatic Monte Carlo~\cite{boldmc}.
Since Taylor expansion in $\lambda$ is subject to the Dyson's collapse argument,
it is clear that some regularization of the theory is required by which we mean
any sequence of field-theoretical approximations to the original model,
controlled by regulator $N$, leading to a diagrammatic series with (at least)
a finite convergence radius. The correct physical result is obtained after
extrapolation of $N$ to infinity.
As an example of such a regularization, we mention a continuous-space
(finite-temperature) system of fermions with truncated single-particle momenta.
Truncation sets the limit on the maximal density of the system,
thereby saving the problem from Dyson's collapse and allowing one to
employ the diagrammatic treatment up to high-order
(see, {\it e.g.}, \cite{Burovski}). While we do not exclude the possibility that
in specific cases the regularization procedure might ultimately be reduced
to a re-summation scheme, we realize that a generic regularization protocol,
in view of the mathematical ambiguity of restoring a function from
an asymptotic series with zero convergence radius, should derive
from the series provenance, {\it i.e.} explicitly depend on the form
of the original theory. We stress that we seek a regularization
which preserves the structure of diagrammatic expansions.

In this Letter, we introduce a class of regularization techniques that introduce
counter-terms to the diagrammatic series. In this approach, the so-called
``sign blessing'', {\it i.e.} sign alternation of the same-order diagrams
(cf. the fermionic case \cite{blessing}) of a regularized
series---as contrasted to the sign-definiteness of the same-order
diagrams of the original theory---is responsible for mutual cancelation
of contributions from the factorial number of diagrams of a given order and
series convergence.

Focusing solely on the convergence properties, it is instructive~\cite{NegeleOrland}
to reduce the theory (\ref{eq:psi_four}) to its  zero-dimensional extreme.
In this case, the field $\psi({\bf r})$
is replaced by a complex number $\psi$ and the partition function reduces to the integral ($x\equiv |\psi|^2$)
\begin{equation}
I(\lambda) = \int_0^{\infty}\!  dx \, {\rm e}^{-x - \lambda x^2}=  \sqrt{\pi\over 4\lambda}\, {\rm e}^{ 1\over 4 \lambda} \, {\rm erfc} \left( {1\over 2 \sqrt{\lambda}} \right)  \, .
\label{eq:integral}
\end{equation}
In order to understand the convergence properties of the perturbative approximations in $\lambda$ for the $\vert \psi \vert^4$ theory, it suffices to do that for the integral $I(\lambda)$.

We start with identifying the problem for the standard Taylor series representation
\begin{equation}
e^{-\lambda x^2} =  \lim_{N \to \infty}
\sum_{k=0}^{N} \frac{(-\lambda)^k}{k!} \; x^{2k}\;,
\label{eq:wrong_sequence0}
\end{equation}
which is that, reversing the order of taking the limit $N\to \infty$ and integrating over $x$,
one obtains a series diverging as $\sim \lambda^k (2k)!/k! $. This is
a hallmark for an asymptotic series with zero convergence radius.
An equivalent observation is that for any finite $N$ the integral is
dominated by values of $x$ where the finite sum is {\it not} providing an accurate
description of the original exponent \cite{Yannick}. Using a different
representation  for the exponential function
\begin{equation}
e^{-\lambda x^2} =  \lim_{N \to \infty} Z_N (x) = \lim_{N \to \infty} \left( 1 - \frac{\lambda x^2}{N}  \right)^N \;,
\label{eq:wrong_sequence}
\end{equation}
also fails for exactly the same reason. It formally produces a sequence of  integrals $I_N(\lambda)$, by replacing  $e^{-\lambda x^2} \to Z_N (x)$, which
takes us {\it away} from the original problem because the dominant contribution to $I_N(\lambda)$ comes from $x \sim N$, as contrasted to the scale
$x\lesssim 1$ of the original integral.
Equation~(\ref{eq:wrong_sequence0}) can be transformed into
Eq.~(\ref{eq:wrong_sequence}) by applying a resummation
technique to the original Taylor series,
\begin{equation}
\left( 1 - \frac{\lambda x^2}{N}  \right)^N \equiv
\sum_{k=0}^{N} \frac{(-\lambda)^k}{k!} \; x^{2k} f(k,N)\;,
\label{resum}
\end{equation}
where the resummation function $f(k,N)= k!C^k_N/N^k $ is such that
$f(k\ll N) \approx 1 $ and decreasing fast for $k\to N$.
Formally, $f$ is supposed to suppress the leading divergence of the original series
which is not the case for Eq.~(\ref{eq:wrong_sequence}). One may seek other
functions $f(k,N)$  rendering the extrapolation procedure $N\to \infty$
meaningful but as far as we know the solution was not found yet, nor is it
clear that it exists. However, as we prove below, the solution does exist
when the series are generalized to include additional counter-terms; namely, within
the
\begin{equation}
e^{-\lambda x^2} = \lim_{N \to \infty} \sum_{k=0} x^k \: f(k,N)
\label{eq:best_sequence}
\end{equation}
representation, where  $f(k,N)$ is restricting summation over $k$
to some polynomial of finite order which scales with $N$.
We require that the expansion is in terms of integer powers of $x$ since otherwise
the construction of Feynman's diagrams (which is our ultimate goal) becomes problematic. To generate the standard
Feynman diagrammatic  expansion  based on Wick's theorem one has to
introduce the exponential
\begin{equation}
Z_N(x) \equiv {\rm e}^{\ln \left[ Z_N(x) \right] }\, ,
\label{eq:exponetialize}
\end{equation}
and expand the logarithm in powers of $x$.

Our solution is based on designing an alternative representation
of the exponential function wich has to satisfy several requirements when
expanded in powers of the coupling parameter. A successful perturbative approach
has to satisfy the conditions of (i) convergence of the $\lambda$-expansion
of $I_N(\lambda)$ for all $\lambda > 0$ at any fixed $N$,
and (ii) meaningful extrapolation to the $N\to \infty$ limit to ensure that
$\lim_{N\to \infty} I_N(\lambda)=I(\lambda)$.
Given that the theory (\ref{eq:psi_four}) features non-analyticity at the second-order phase transition point, one might think that the condition (i) should be weakened to sufficiently small $\lambda$'s. Nevertheless, we consciously require convergence everywhere
to guarantee that the divergence of the $\lambda$-expansion of the field-theoretical counterpart of  $I_N(\lambda)$ is exclusively due to the phase transition
in the macroscopic limit;  the  convergence radius remaining infinite for any finite-size system (qualitatively analogous to our zero-dimensional model).

Let us introduce the parameterization
\begin{equation}
y = u  x^{2/m}, \qquad \quad u = \left( \lambda /N \right)^{1/m} \, ,
\label{eq:parameterization}
\end{equation}
with $m$ an integer number, and note that the most severe problem with
Eq.~(\ref{eq:wrong_sequence}) can be overcome by seeking a
solution in the form
\begin{equation}
{\rm e}^{-\lambda x^2} = \lim_{N \to \infty} \left[ f(y) \right]^N \;,
\label{eq:aga}
\end{equation}
with the function $f(.)$ being bounded, $0\le  f(.) \le 1$, along the real axis
of $y$, and having an expansion
\begin{equation}
f(y) = 1 - y^m + \dots \, .
\label{eq:requirement}
\end{equation}
The terms that are of a higher exponent than $m$ serve as counter terms compared to Eq.~(\ref{eq:wrong_sequence}). These requirements guarantee that the series of
integrals $I_N(\lambda)$ converge to the right answer, but this does not solve yet
the problem with perturbative expansions unless we also require that
the integral
\begin{equation}
I_N(\lambda) = \int_0^{\infty}\!  dx \, {\rm e}^{-x }\left[ f(y(x)) \right]^N
\label{eq:integral_N}
\end{equation}
is convergent for any complex value of $u$ [see Eq.~(\ref{eq:parameterization})], which is the necessary condition for $I_N(\lambda)$ to be expandable in the convergent Taylor series in powers of $u$. It is this condition which forces one to consider $m>1$.

It turns out that there are infinitely many functions corresponding to $m=4$ ($m=4$ is in practice the obvious choice),  all satisfying the above-formulated requirements. Here we present one of them
which is based on Bessel functions $J_n$:
\begin{equation}
f(y) =J_0(2z)+2J_2(2z)+\frac{5}{3}J_4(2z) \, ,
\label{eq:bessel_expansion}
\end{equation}
where $z=(72)^{1/4}\;y $ and the explicit values of coefficients follow from the
requirement (\ref{eq:requirement}). With regards to all conditions we find: \\
 - For any positive  $u$ the condition  $0 \le f(.) \le 1$
means that all integrals (\ref{eq:integral_N}) are dominated by
the $x\lesssim 1$ region and thus guaranteeing that the condition (ii) is satisfied. \\
- $[f(y)]^N$ is an entire function with infinite convergence radius.\\
- The integrals (\ref{eq:integral_N}) converge for any complex $u$ since the strongest divergence at $x\to \infty$ (occurring at imaginary $u$) is exponential with the exponent $\propto \sqrt{x}$ and thus not dangerous in view of the ${\rm e}^{-x}$ factor; the condition (i) is thus satisfied. \\
 - the Taylor expansion of $f(y)$ is in even powers of the argument.
Hence, the requirement of having only integer powers of $x$ is also satisfied since $y^2=u^2x$.

\begin{figure}[h]
\begin{center}
\includegraphics[width=1\columnwidth]{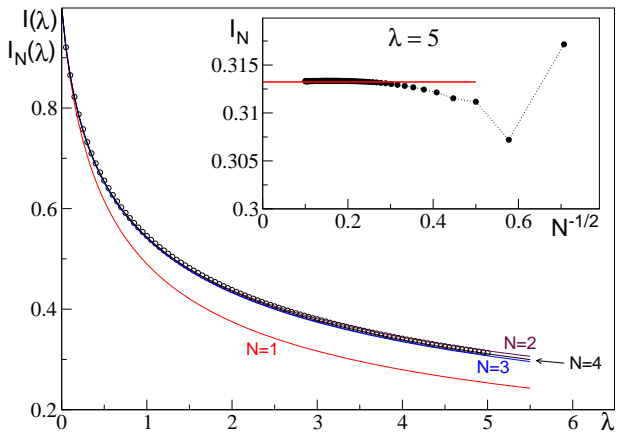}
\caption{(color online) Convergence properties of the sequence $I_N(\lambda)$
with $f(y)$ defined in Eq.~(\ref{eq:bessel_expansion}). The result for $I(\lambda)$ is shown with circles.
The convergence with $N$
for $\lambda=5$ is shown in the inset. }
\label{fig:convergence_bessel}
\end{center}
\end{figure}
\begin{figure}[h]
\begin{center}
\includegraphics[width=1\columnwidth]{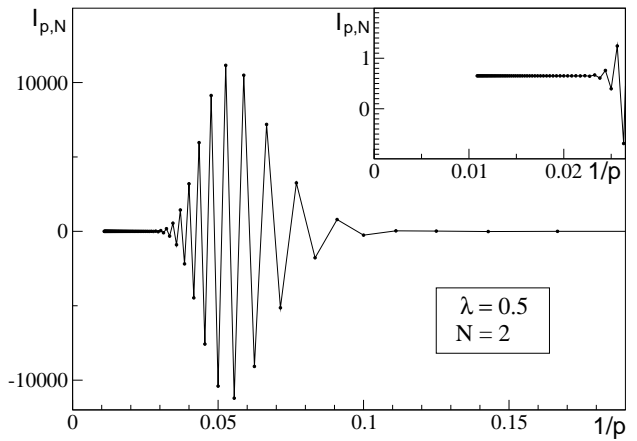}
\caption{Plotted are integrals $I_{p,N} = \int dx \, e^{-x} Z_{p,N}(x)$
where $Z_{p,N}$ is the Taylor expansion of $[f(y)]^N$ up to order
$x^p$, for $\lambda=0.5$. In this example $N=2$ which guarantees
that the converged answer is accurate down to one percent accuracy.}
\label{fig:taylor}
\end{center}
\end{figure}
\begin{figure}[h]
\begin{center}
\includegraphics[width=1\columnwidth]{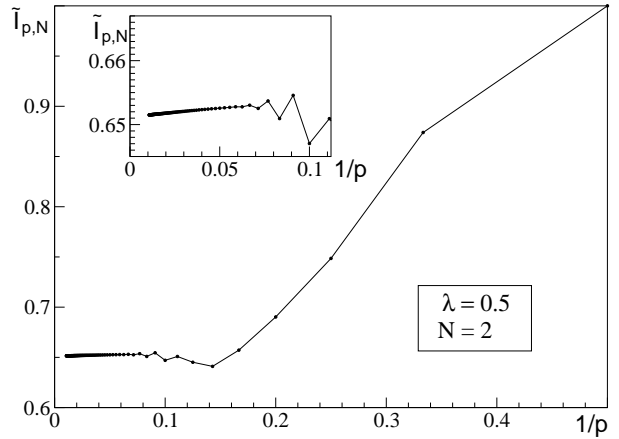}
\caption{Integrals $\tilde{I}_{p,N} = \int dx e^{-x} \tilde{Z}_{p,N}(x)$
based on the same series as in Fig.~\ref{fig:taylor} resummed using
the near-Gaussian function, see text. The elimination of the severe sign-problem and
the radical improvement of convergence properties suggests that
polynomial size and optimization should be explored.}
\label{fig:taylorG}
\end{center}
\end{figure}

The convergence with $N$ of the sequence of integrals (\ref{eq:integral_N})
is illustrated  in Fig.~\ref{fig:convergence_bessel}. Remarkably, $N=4$ already
produces results which are accurate at the 1\% level up to $\lambda=5$. The
ultimate convergence to the exact solution is a polynomial function of
$N^{-1/2}$ and can be easily extrapolated from finite $N$.
A peculiar feature of the arising theory is
that perturbative expansion is in terms of $u$ which is a
fractional power of $\lambda$.

In order to comply with Feynman's diagrammatic rules we need to
study integrals over $x$ for Taylor expansions
$Z_N(x)=\sum_{k=0}^{\infty}x^k C_k$ done up to order $p$.
The implications for that are shown in Fig.~\ref{fig:taylor}
where we plot the final result when $Z_N(x)$ is replaced with
$Z_{p,N}(x) = \sum_{k=0}^p x^k C_k$.
At first, the impression is that a severe sign-problem
and a very high-order expansion is the price for obtaining a set
of approximations using series with infinite convergence radius.
When we take stronger $\lambda$ or larger $N$, the fluctuations that
need to cancel against each other become larger, and the minimum Taylor expansion order increases. However, when the original series are further modified using
resummation technique based on the near-Gaussian function
\begin{equation}
Z_{p,N}(x) \to \tilde{Z}_{p,N}(x) = \! \sum_{k=0}^p x^k C_k e^{-k^s/p}  , ~~~ s=2.05 , ~
\label{eq:Gaussresumm}
\end{equation}
we obtain results shown in Fig.~\ref{fig:taylorG}. Amazingly, the sign problem
between the expansion orders is essentially eliminated and accurate results
can be obtained already for relatively small expansion orders $\sim 5$,
making the entire scheme
a viable solution to the regularization of diagrammatic expansions with
zero-convergence radius (increasing $N$ does not change the picture substantially).

At this point we note that series convergence for fixed $N$ allows us to truncate
the series at some order $p_{*}(N)$ such that the difference between $I_{p_{*},N} = \int dx\,  e^{-x} Z_{p_{*},N}(x)$
and $I_N$ is smaller than
$\vert I_N - I(\lambda) \vert$.
This would account for the regularization technique in the
form (\ref{eq:best_sequence}) when the set of approximations to
the exponential function is always in the form
of a {\it finite-order} polynomial in $x$. Clearly, there is a lot of room
for optimizing the form of such polynomials, and the success of the
near-Gaussian resummation, which in essence performs such an optimization,
proves the point---radical improvements in efficiency are possible along these
lines even with relatively small $p_{*}$.
One of such approaches (tested for our model with quite satisfactory results) is based on polynomials that reproduce
exactly the first $p_{*}$ moments of the distribution $e^{-x-\lambda x^2}$.

To deal with a generic {\it non-local} coupling between fields
$ e^{-\lambda \vert \psi_1 \vert^2 \vert \psi_2 \vert^2} \equiv e^{-\lambda x_1 x_2}$
we propose to start with the following regularization (later on it can be
further modified/optimized)
\begin{equation}
e^{-\lambda x_1x_2} = \lim_{N \to \infty}
\left[ g(y_1)g(y_2) - \frac{\lambda x_1x_2}{N} f(y_1)f(y_2) \right]^N ,
\label{eq:generic}
\end{equation}
where $y_i = u' x_i^{1/2}$, and $u'=(1/N^a)^{1/2}$ with the exponent $a<1/2$.
The entire functions $g(y)$ and $f(y)$
have to satisfy criteria similar to those mentioned above: be even functions of
$y$, decay on the real axis, and diverge not faster then $e^{\vert y \vert }$
in the complex plane. These requirements are satisfied if $f$ and $g$ are
based on linear combinations of Bessel functions.
We also demand that $f(0)=g(0)=1$, the Taylor series expansion of
$g$ starts from an order $y^{2s}$ with
$as>1$, and the modulus of expression in parentheses is
smaller than unity. This also constitutes no problem using proper linear
superpositions of $J_n$. The resulting regularization is
guaranteed to work and, most importantly, it can be generalized to deal with
arbitrary interaction potentials (two-body, three-body, etc.)
because the $f$ and $g$ functions are defined in terms of local fields only.
We had to proceed in this way instead of repeating the solution of the
single-field problem because the expansion in
$y^2 \propto \vert \psi_1 \vert \vert \psi_2 \vert $ does not allow
the diagrammatic technique, contrary to $y_i^2 \propto \vert \psi_i \vert^2$.

The other important difference between Eqs.~(\ref{eq:generic}) and
(\ref{eq:aga}) is that the  $y$ variables in Eq.~(\ref{eq:aga})
do not contain the parameter $\lambda$, and
the diagrammatic expansion goes thus in integer powers of $\lambda$ and $u'^2$.
We believe that with an appropriate choice of functions it is possible to proceed
with $a=1/2$ and $u'=(\lambda/N)^{1/4}$, just as before. This approach has the
``advantage'' of being based on a single parameter expansion.
Likewise, one may apply (\ref{eq:generic}) with $a<1/2$ and an
$\lambda$-independent parameter $u'$ to deal with the single-site
regularization.

In conclusion, diagrammatic series with zero convergence radius can be dealt with by
approximating the original interaction exponential with an order-$p$ polynomial in the integer powers of the original
interaction and regularizing counter-terms. The action of the regularized theory is  obtained by exponentiating the polynomial.
The accuracy of the approximation is controlled by the parameter $p$.
While the polynomial has to satisfy a number of requirements guaranteeing, in particular, that the sequence of approximate theories approaches the original
one at $p\to \infty$ (counterintuitively, this property
requires special care), there is still a continuum of possible choices, with some characteristic examples tested  on the basis of zero-dimensional $|\psi|^4$ model.
We believe that the outlined approach---especially in the context of Diagrammatic Monte Carlo---opens up an opportunity to utilize Feynman's diagrams as a generic tool to address
strongly correlated classical- and quantum-field systems.

We acknowledge valuable and stimulating discussions with M. Ogilvie and Y. Meurice
during our stay at the Aspen Center for Physics. This work was supported by the
Swiss National Science foundation, the National Science Foundation
grant PHY-0653183, and by a grant from the Army Research Office
with funding from the DARPA OLE program.


\begin{thebibliography}{99}
\bibitem{Dyson52} F. J. Dyson, Phys. Rev. {\bf 85}, 631 (1952).
\bibitem{NegeleOrland} J. W. Negele and H. Orland, Quantum Many-Particle Systems, Westview Press, Boulder, 1998.
\bibitem{boldmc} N. V. Prokof'ev and B. V. Svistunov, Phys. Rev. Lett. {\bf 99}, 250201 (2007).
\bibitem{Burovski} E. Burovski, E. Kozik, N. Prokof'ev, B. Svistunov, and M.
Troyer,  Phys. Rev. Lett. {\bf 101}, 090402 (2008).
\bibitem{blessing}  N. Prokof'ev and  B. Svistunov,  Phys. Rev. B {\bf 77}, 020408 (2008);
K. Van Houcke, E. Kozik, N. Prokof'ev, and B. Svistunov,
{\it Diagrammatic Monte Carlo},  [in Computer Simulation Studies in Condensed Matter Physics XXI, Eds.
D.P. Landau, S.P. Lewis, and H.B. Schuttler (Springer Verlag,
Heidelberg, Berlin 2008)]; E. Kozik, K. Van Houcke, E. Gull, L. Pollet, N. Prokof'ev, B. Svistunov, and M. Troyer,
 EPL {\bf 90}, 10004 (2010).
 \bibitem{Yannick} The idea of our approach to series regularization
was inspired by Y. Meurice's discussion of how naive regularization methods fail.

\end{thebibliography}
\end{document}